\documentclass[aps,prl,twocolumn,groupedaddress,showpacs]{revtex4}
\usepackage{graphicx}

\newcommand{\delete}[1]{}

\newcommand{\be}{\begin{equation}}
\newcommand{\ee}{\end{equation}}

\def\beq{\begin{equation}}
\def\eeq{\end{equation}}
\def\bea{\begin{eqnarray}}
\def\eea{\end{eqnarray}}
\def\ba{\begin{array}}
\def\ea{\end{array}}

\begin{document}


\title{Zeptonewton force sensing with nanospheres in an optical lattice}
\author{Gambhir Ranjit, Mark Cunningham, Kirsten Casey, Andrew A. Geraci}
\email[]{ageraci@unr.edu}
\affiliation{Department of Physics, University of Nevada, Reno, Reno NV, USA}


\date{\today}

\begin{abstract}
Optically trapped nanospheres in high-vaccum experience little friction and hence are promising for ultra-sensitive force detection. Here we demonstrate measurement times exceeding $10^5$ seconds and zeptonewton force sensitivity with laser-cooled silica nanospheres trapped in an optical lattice. The sensitivity achieved exceeds that of conventional room-temperature solid-state force sensors by over an order of magnitude, and enables a variety of applications including electric field sensing, inertial sensing, and gravimetry. The particle is confined at the anti-nodes of the optical standing wave, and by studying the motion of a particle which has been moved to an adjacent trapping site, the known spacing of the anti-nodes can be used to calibrate the displacement spectrum of the particle.  Finally, we study the dependence of the trap stability and lifetime on the laser intensity and gas pressure, and examine the heating rate of the particle in vacuum without feedback cooling.

\end{abstract}


\maketitle


Sub-attonewton force sensing facilitates a variety of applications including magnetic resonance force microscopy \cite{rugar2}, tests of gravitational physics at short range \cite{stanford08,beadprl}, investigations of surface forces including the Casimir effect \cite{casimir}, as well as inertial sensing \cite{accelerometer}. State-of-the-art resonant solid state mechanical sensors such as micro-cantilevers, nano-membranes, and nanotubes typically operate in a cryogenic environment to improve their thermal-noise limited force sensitivity. Room-temperature solid-state sensors have achieved sensitivity in the $\sim 10-100$ aN$/{\rm{Hz}}^{1/2}$ range \cite{roomtemp1,roomtemp2,roomtemp3,roomtemp4,roomtemp5}, while cryogenic nanotube mechanical oscillators have recently achieved $\sim 10$ zN$/{\rm{Hz}}^{1/2}$ \cite{nanotube}. The excellent environmental decoupling of optically levitated mechanical systems \cite{changsphere,virus,raizen,rochester,aspelmeyercavity,quidant13,barkeriontrap} in high vacuum can allow such systems to achieve similar or better force sensitivity at room temperature \cite{levreview,quidant13,rodenburg}. However, a challenge has been the optical confinement of such particles under high vacuum \cite{rodenburg,ashkin3,barkerheating,atherton2015}, in particular in standing-wave optical traps \cite{aspelmeyercavity,quidant15}.

In this paper we describe robust optical trapping of $300$ nm silica nanospheres in an optical lattice at high vacuum, where particles can be trapped indefinitely over several days. The optical potential allows the particle to be confined in a number of possible trapping sites. By perturbing the system with a laser, we are able to transfer the particle between different trap anti-nodes, which shows promise for sensing experiments where the particle position must be adjusted and controlled precisely \cite{beadprl}.  By studying the motion of a particle which has been moved to an adjacent trapping site, the known spacing of the lattice anti-nodes can also serve as a ruler to calibrate the displacement spectrum of the particle.  While electric fields can be used to calibrate the force sensitivity of charged microspheres \cite{atherton2015,millicharge}, the standing wave method can be a useful calibration tool for neutral objects, which are applicable for a variety of experiments where charge can produce unwanted backgrounds. We find that for a charged particle the standing-wave method produces results consistent with the electric field method.

Using active-feedback laser cooling in three dimensions, we demonstrate cooling of the center of mass motion to $\sim 400$ mK at a pressure of $5 \times 10^{-6}$ Torr, resulting in a force sensitivity of $1.6$ aN$/{\rm{Hz}}^{1/2}$. The system permits time-averaged measurements over long integration times, and we demonstrate force sensing at the 6 zN level. Due to the reduced particle size and improved imaging and feedback cooling, these results are more than two orders of magnitude more sensitive that those previously reported by our group using $3$ $\mu$m particles in a dual-beam optical dipole trap \cite{atherton2015}.

Finally, we study the dependence of the trap stability and lifetime on laser intensity and background gas pressure, and measure the heating rate of the particle in high vacuum in the absence of optical feedback cooling. We find stable trapping for a range of intensities that are limited by the trapping depth on one hand and the internal heating of the particle on the other.

In addition to force sensing applications, stable optically trapped nanospheres at high vacuum are also promising for quantum information science \cite{changsphere,virus}, tests of classical and quantum thermodynamics \cite{barkerheating}, testing quantum superpositions \cite{oriol, NVcenters1,NVcenters2}, quantum opto-mechanics with hybrid systems \cite{ranjitpra}, matter wave interferometers \cite{barkerAI,tongcangdiamond,arndt,ulbricht,hartandy}, and gravitational wave detection \cite{GWprl}.

\begin{figure}
\includegraphics[width=0.9\linewidth]{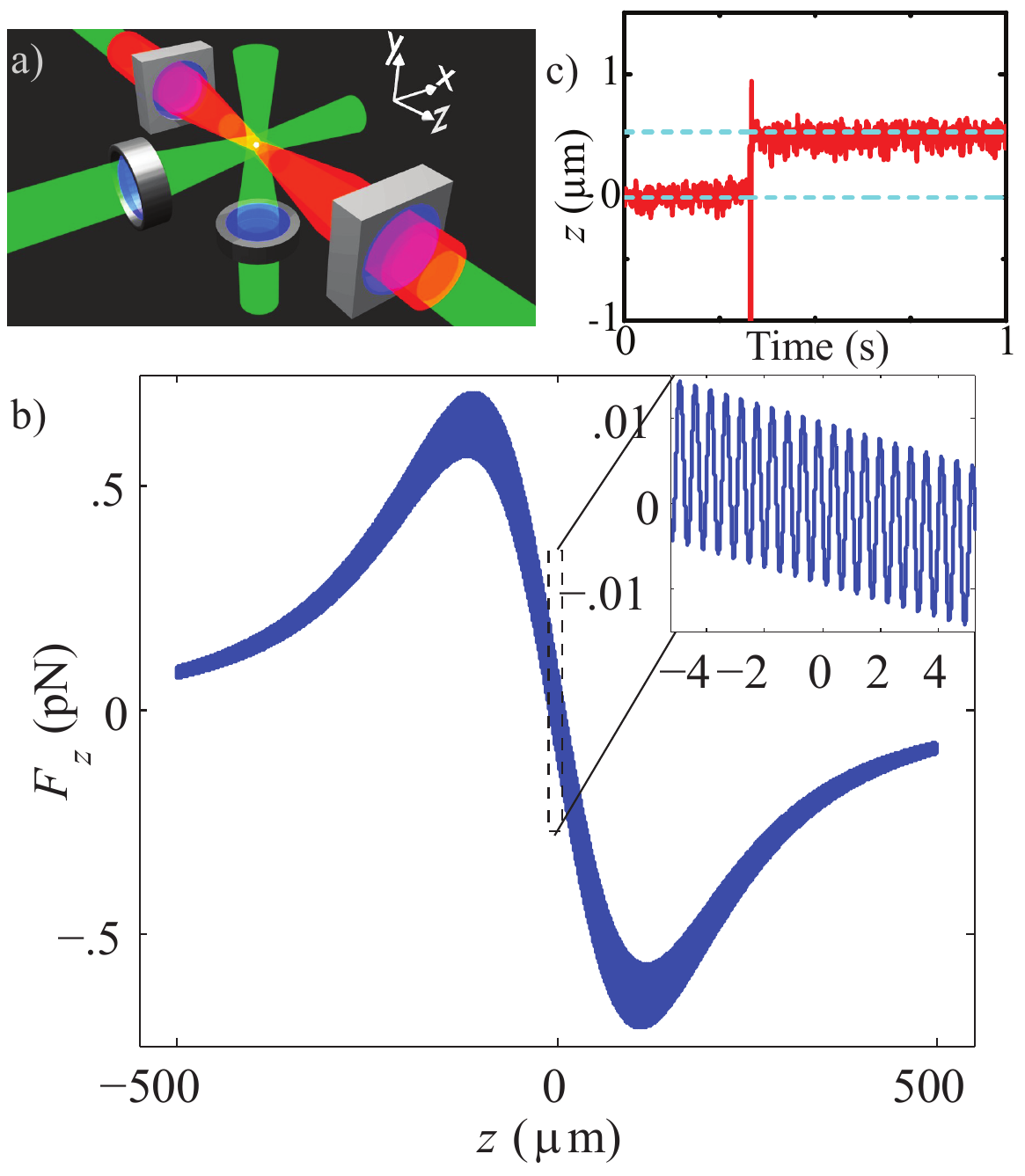}
\caption{\label{fig:standingwave}(Color Online) (a) A standing-wave trap for 300 nm beads is formed using counter-propagating 1064 nm laser beams focused at nearly the same spatial location. Active feedback cooling is performed using 780 nm lasers (shown as green) in 3 dimensions. (b) Calculated optical force along the $z-$axis assuming total power of $2.2$ W, waist of $8$ $\mu$m, and a $0.2\%$ intensity modulation due to interference from the counter-propagating beam, corresponding well with the measured trap frequencies. (c) Time-trace for $1$ s of particle motion in the axial direction at $P=2$ Torr. When subject to an applied sinusoidal optical force, the particle hops to an adjacent trapping site as a result of the perturbation. Dotted lines indicate expected antinode spacing.} 
\end{figure}

\emph{Experimental Setup.}
A schematic of the experimental setup is shown in Fig. \ref{fig:standingwave}.  A $300$ nm fused silica sphere is trapped using two equal-power counter-propagating beams formed by splitting a $1064$ nm laser beam with a polarizing cube beam splitter. The beam foci are offset axially by $75$ $\mu$m. 
The trap is initially operated with a total power of $2.2$ W and a waist size of approximately $8$ $\mu$m, and the trap is loaded by vibrating a glass substrate to aerosolize beads under $5-10$ Torr of $N_2$ gas, which provides sufficient damping to slow and capture the particles.  More detail of the vacuum system has been previously described in Ref. \cite{atherton2015}.

The polarizing cube beam splitter transmits approximately $1.5$ percent of the p-polarized laser power along the s-beam path due to imperfect polarization separation. This p-polarized component can interfere with the anti-parallel p-polarized beam to create a standing wave potential, as illustrated in Fig. \ref{fig:standingwave}b. The optical potential results from the superposition of the scattering and dipole forces from the beams and includes a modulation produced by the interference. The intensity modulation depends on the coherence length of the laser as well as the purity of the beam polarizations.

The position of the nanosphere is measured by imaging the scattered light from the nanosphere
onto two quadrant photodetectors (QPDs). We define the ``axial'' or $z-$ axis in the direction of the dipole trap beams, and the ``horizontal'' or $x-$ axis is perpendicular to both the vertical and axial axes.  The axial-horizontal (vertical) motion is measured using QPD 1 (2). 
The position signals from the QPDs are phase shifted by $90$ degrees to provide a signal proportional to the bead's instantaneous velocity using either a derivative or phase shifter circuit.  The phase shifted signals are used to adjust the RF amplitude of three acoustic optical modulators (AOMs), which modulate the intensity of a $780$ nm laser beam to provide a velocity-dependent optical damping force in each direction. Such feedback has proven necessary for maintaining the particle in the trap while pumping to high-vacuum. The feedback light is focused onto the sphere using a lens outside of the vacuum chamber in the horizontal direction, one of the dipole trap lenses for the axial direction, and an in-vacuum lens for the vertical direction.

Prior to pumping to high vacuum, the center-of-mass temperature as derived from the position spectrum of the beads is largely independent of pressure and trap laser power for sufficiently high pressure and sufficiently low laser intensity. We can thus assume the bead is in thermal equilibrium with the background gas at and above $2$ Torr.  This allows us to determine a scale factor to convert the quadrant photodetector voltage into a displacement.  From this conversion factor we can deduce the force sensitivity of the bead at lower vacuum conditions. As a check of the scale factor, the bead can be transferred between adjacent trapping sites by applying a perturbation with a laser. In this case we utilize the feedback cooling laser in a driving mode.  In Fig. \ref{fig:standingwave}c we show the time trace of a bead subject to a perturbation which causes it to transition between adjacent trapping sites. A calibration is made possible using the half-wavelength spacing of the trap antinodes, along the axial direction of the trap. From the fit to thermal spectra, the measured displacement of this transition is $514 \pm 43$ nm, in reasonable agreement with the expected value of $532$ nm.

\emph{Force Measurement.}
At high vacuum, time-averaged sub-aN force measurements can be performed.
The minimum force detectable for a harmonic oscillator in thermal equilibrium with a bath at temperature $T$ is
\begin{equation} \label{Fmin1}
F_{\rm{min}}=S_{F}^{1/2} b^{1/2}=\sqrt{\frac{4 k_{B} T b k}{\omega_{0}Q}}
\end{equation}
where $b$ is the measurement bandwidth, $S_{F}^{1/2}$ is the thermal-noise force spectral density , $k$ is the spring constant of the oscillator, $k_{B}$ is Boltzmann's constant, $w_{0}$ is the resonance frequency, and $Q$ is the quality factor. In the absence of laser cooling, Eq. \ref{Fmin1} can be written for a nanosphere as $F_{\rm{min}}=\sqrt{4 k_{B} T m\Gamma_M b}$
where $\Gamma_M=16P/(\pi\rho v r)$ is the damping coefficient of the surrounding gas, $v$ is the mean speed of the gas, $\textit{m}$ is the mass of the sphere, $\rho$ is its density, $\textit{r}$ is its radius, and $P$ is the pressure.  For a sphere cooled with laser feedback cooling, the temperature in Eq. \ref{Fmin1} becomes $T_{\rm{eff}}$ and the damping rate $\Gamma_{\rm{eff}}$ includes the effect of the cooling laser.

\begin{figure}
\includegraphics[width=\linewidth]{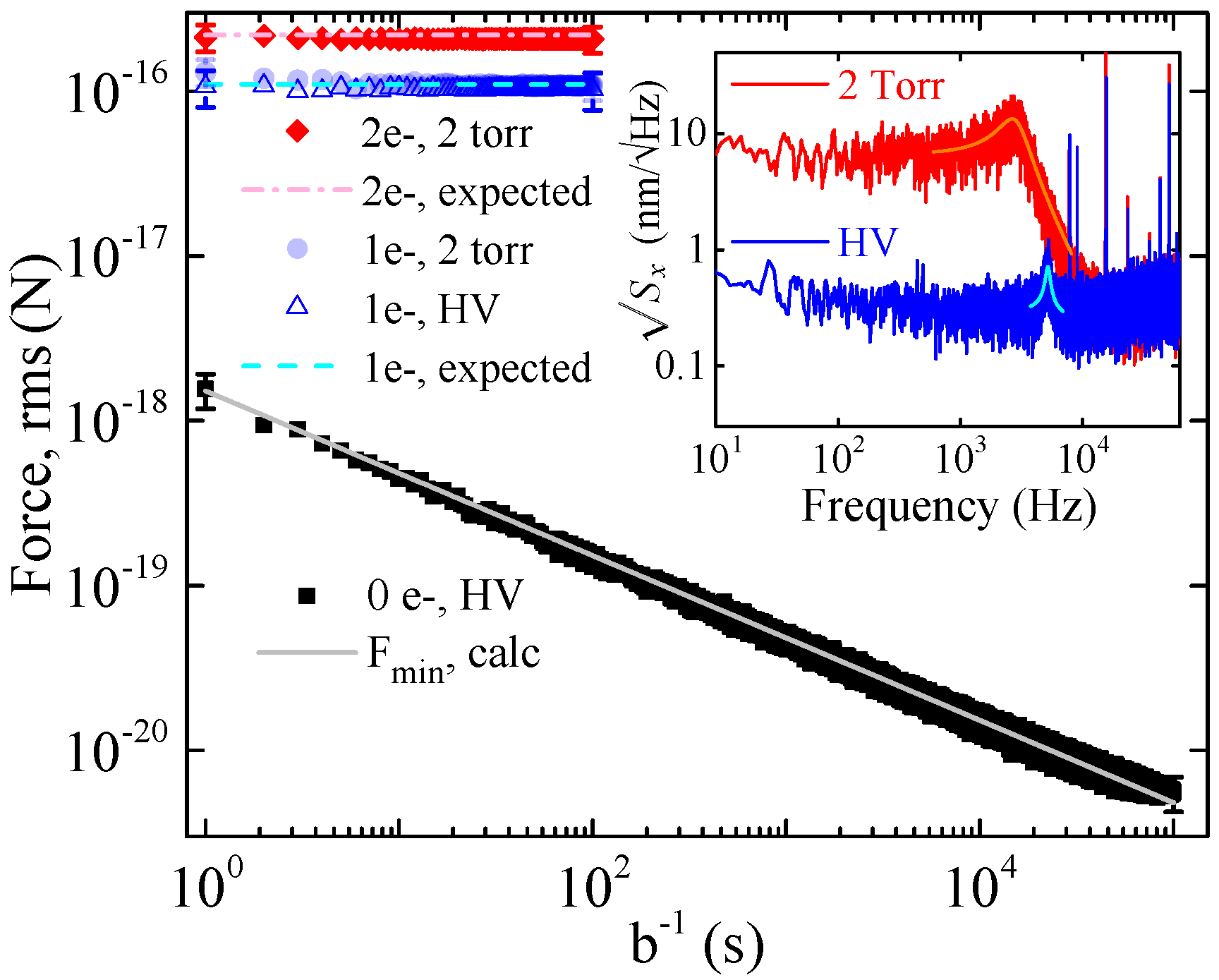}
\caption{\label{fig:thnoise}(Color Online) Measured force on a bead as a function of averaging time at 2 Torr and $5 \times 10^{-6}$ Torr (HV) for charged and uncharged beads, while driving with a sinusoidally varying electric field of 1 kV/m. (inset) Measured $x-$ displacement spectrum of a $300$ nm sphere at $2$ Torr and HV with feedback cooling applied. Lorentzian fits indicate cooling to 460 mK at HV.}
\end{figure}

We perform force measurements in the $x-$direction. Data for the bead position and a reference signal (typically at 9 kHz) are recorded with a sampling rate of 125 kHz. Fig. \ref{fig:thnoise} shows a typical displacement spectral density in the $x-$direction of a bead held at low vacuum of $2$ Torr with no feedback cooling applied, and a spectrum at high vacuum (HV) of $5 \times 10^{-6}$ Torr with feedback cooling. At 2 Torr we observe an \emph{x-}resonant frequency of 2830 Hz and gas damping rate of approximately 1.4 kHz. In the orthogonal directions $(y-,z-)$ (not shown) resonance frequencies of $(3410,7300)$ Hz are observed, respectively. At HV, a lorentzian fit to the data reveals cooling of the center of mass motion to $460 \pm 60$ mK, with a damping rate of $460 \pm 49$ Hz in the $x-$direction.  CM motion in the $y-$ and $z-$directions are cooled to temperatures of $610 \pm 190$ mK and $7.9 \pm 3$ K, with damping rates of approximately 1.3 kHz and 1 kHz, respectively. The frequencies of the peaks are shifted when feedback cooling is applied due to the optical spring effect that occurs if the feedback phase is not precisely 90 degrees. The force sensitivity in the $x-$direction corresponds to $S_{F,x}^{1/2}=1.63\pm0.37$ aN$/{\rm{Hz}}^{1/2}$, with the error dominated by the uncertainty in the particle size. The lowest attainable temperature appears to be limited by noise in the QPD imaging electronics and trapping laser. The expected sensitivity at this pressure would be approximately $\sim 10$ times better in the absence of laser noise and cross-talk between feedback channels. 


 In the absence of an applied force, we expect the signal due to thermal noise to average down as $b^{1/2}$. This behavior is shown in Fig. \ref{fig:thnoise} for averaging times exceeding $10^5$ seconds.  Force sensing at the level of $5.8 \pm 1.3$ zN is achievable at this timescale. Also shown is the calculated $F_{\rm{min}}$ using the measured parameters for $T_{\rm{eff}}$, $\omega_0$, and $\Gamma_{\rm{eff}}$, which agree with measured data within uncertainty. We find that approximately $90 \%$ of the beads trapped have zero electric charge; the remaining beads tend to have only $1$ or $2$ excess electrons.  Data are shown for charged $(1e-,2e-)$ and uncharged beads in Fig. \ref{fig:thnoise} for a known applied electric field. The expected force for a charge of 1 (2) electrons is shown as a dotted line in Fig. \ref{fig:thnoise}.  An independent calibration can be achieved by comparing the spectra of the beads after they have been transported to adjacent trapping sites in the optical lattice, as discussed previously. The determined calibration factors are consistent in each case within experimental uncertainty.

\begin{figure}
\includegraphics[width=\linewidth]{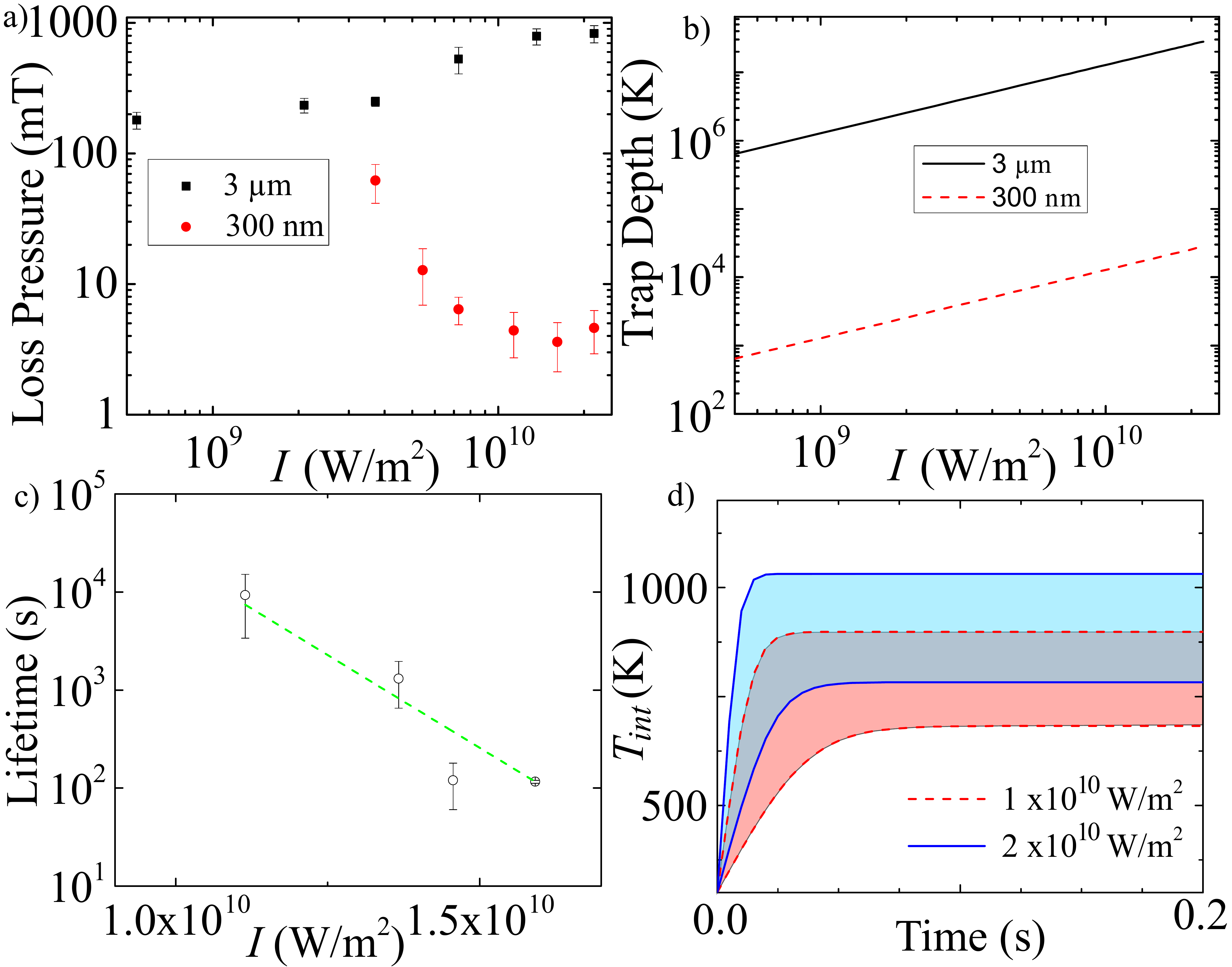}
\caption{\label{fig:beadLoss}(Color Online) (a) Mean pressure at which beads are lost for various laser trapping intensities, with no feedback-cooling applied. Statistics are shown for 30 beads of each size. (b) Calculated trap depth for $300$ nm and $3$ $\mu$m beads. (c) Trap lifetime at high-vacuum versus laser intensity for higher intensity trapping, along with linear-fit. (d) Expected internal temperature rise in a 300 nm sphere versus time for laser intensities of $10^{10}$ and $2 \times 10^{10}$ W/m$^2$, respectively. The shaded bands are determined by varying $\epsilon_2=2.5 \times 10^{-7}$ to $\epsilon_2=1.0 \times 10^6$.}
\end{figure}

\emph{Trap Stability and Lifetime.}
\label{stability}
In the absence of applied feedback cooling, the particle is lost from the trap as the pressure is dropped below the 10 mTorr range. Fig {\ref{fig:beadLoss}}a illustrates statistics for the typical trap loss pressure for beads without feedback cooling applied, as a function of trapping laser intensity, along with previous data obtained for $3$ $\mu$m diameter beads \cite{atherton2015}. 
Following a similar analysis to that presented in Ref. \cite{atherton2015}, we find that radiometric forces may also be a likely loss mechanism for the smaller beads. The expected temperature gradient across the sphere is significantly reduced for the 300 nm sphere however, consistent with the lower loss pressures.  Once HV is attained, we can reduce the optical feedback cooling rate by over an order of magnitude compared with what is used while pumping from 2 Torr to HV, and maintain the trap stability. This suggests that gas collisions play a role in the loss mechanism around $\sim 10$ mTorr. While larger beads tend to be lost at higher pressures for increasing intensity, the 300 nm beads tend to get lost at higher pressures for decreasing intensity. This difference may be due to the reduced trap depth for the smaller particles.

The trap lifetime at high vacuum at intensities around $10^{10}$ W/m$^2$ is typically indefinite over several days, however at higher intensity we notice an exponential reduction of lifetime with increasing laser power, as shown in Fig. \ref{fig:beadLoss}c. The estimated timescale to reach thermal equilibrium in each case is less than $1$ s, as shown in Fig. \ref{fig:beadLoss}d, despite lifetimes ranging from minutes to a few hours. Here we consider a range of possible values for the imaginary permittivity $\epsilon_2$, varying from the bulk silica value $\epsilon_2=2.5 \times 10^{-7}$ \cite{silicaloss} up to $\epsilon_2=10^{-6}$, an upper bound we infer from holding particles for several seconds at intensities above $2 \times 10^{10}$ W/m$^2$ without particle evaporation or loss.  The exact loss mechanism shown in Fig. \ref{fig:beadLoss}c is uncertain. A process whereby the particle may undergo annealing or a glass-crystalline transition after remaining at an elevated temperature for a significant time could be responsible for loss if the new phase has higher absorption or if the bead experiences a kick due to a sudden change in density, size, or refractive index.  Annealing is reported for certain forms of silica at temperatures as low as $500$ K over $30$ minute timescales \cite{glasstransitions}.

\begin{figure}
\includegraphics[width=0.9\linewidth]{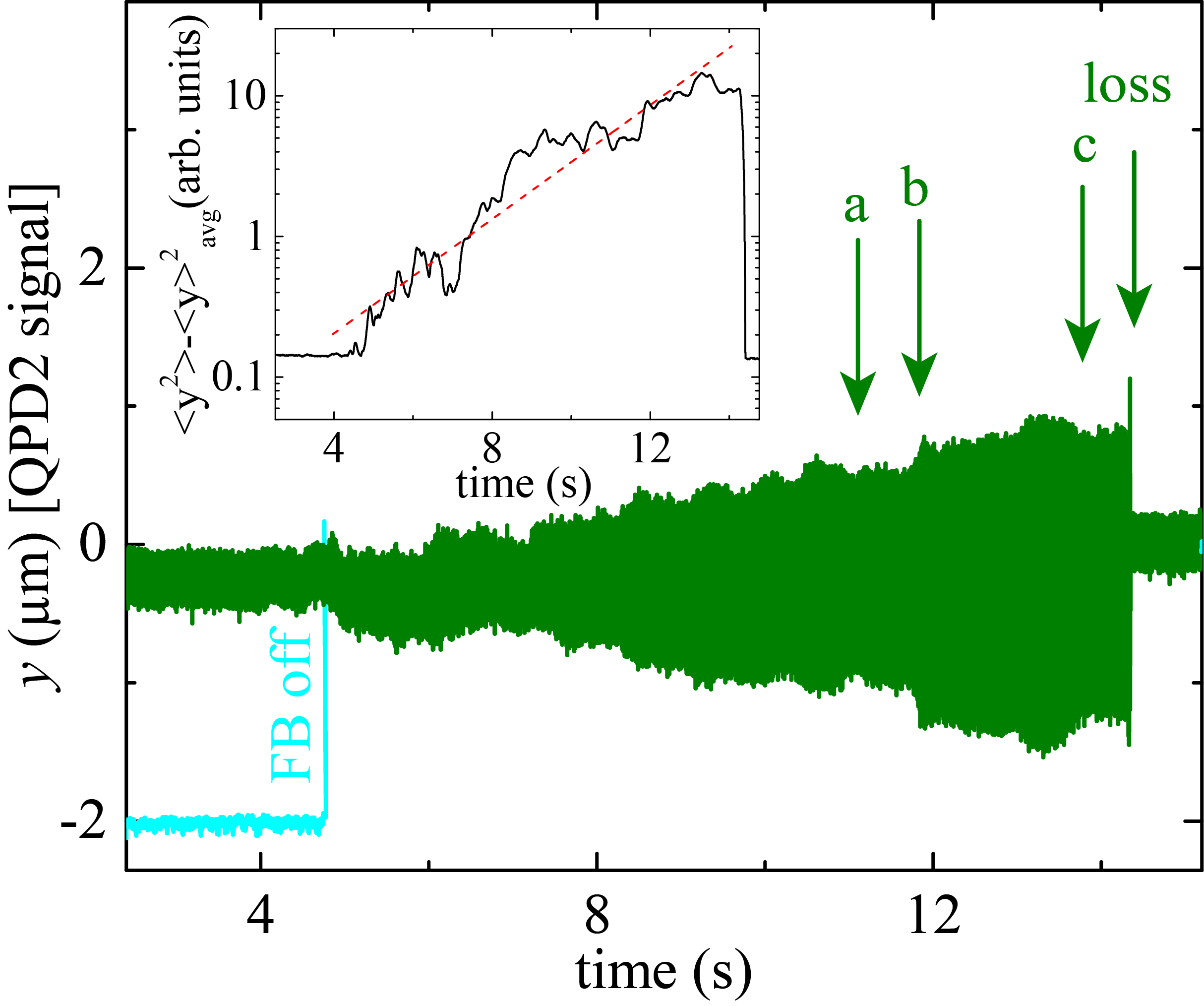}
\caption{\label{fig:jumps}(Color Online) Measured y-displacement signal of the bead as optical feedback cooling is turned off under high vacuum conditions. ``FB off'' indicates the time of blocking feedback lasers. Arrows represent bead hopping to (a) and returning from (b) and then returning to (c) an adjacent lattice site (in the z-direction). (inset) Variance signal with averaged moving window of 0.12 s and fit to exponential growth before bead loss.}
\end{figure}

To evaluate the trap heating rate we study the motion in high vacuum after the optical feedback cooling has been turned off (Fig. \ref{fig:jumps}). As the amplitude increases we observe discrete transitions of the bead between adjacent lattice sites before the bead is ejected from the trap. From an exponential fit to the envelope of the averaged variance we infer a heating timescale of $\sim 200$ mHz.  The expected heating rate from gas collisions $\Gamma_M$ is approximately $3.4$ mHz, while the calculated photon-recoil heating rate is approximately $100 \times$ slower. Thus an additional heating mechanism is present which can include contributions from laser noise and non-conservative scattering forces \cite{nonconserv1}. Given the measured cooling rates, the achieved minimal temperatures are roughly as expected with the measured heating rate.

\emph{Discussion.} We have demonstrated zN level force sensing using nanospheres in a standing-wave optical trap with integrated measurement times exceeding $10^5$s. 
The known spacing of the lattice anti-nodes can serve as a ruler to calibrate the displacement spectrum of uncharged particles, which are often desirable in precision measurements, including Casimir force studies and gravitational experiments.
The ultimate force sensitivity is limited theoretically at low pressure due to photon recoil heating \cite{levreview} and measurement backaction noise \cite{rodenburg}.  
In practice our sensitivity is limited by technical laser noise and noise in the imaging electronics. 
By using cavity assisted displacement readout or by improving the numerical aperture of the imaging system, along with using near shot-noise limited detection electronics and improved laser intensity stabilization, $\sim 100 \times$ colder center-of-mass temperatures and an order of magnitude better force sensitivity should be possible at these pressures.  In this system it should also be possible to study multiple trapped particles, allowing investigations of interaction effects such as optical binding \cite{opticalbinding}.

\emph{Acknowledgements.}
We thank D. P. Atherton and J. Stutz for experimental assistance at the early stages of this work. We thank W. P. Arnott, T. Li, N. Kiesel, and M. Bhattacharya for useful discussions. This work is supported by grants NSF-PHY 1205994, 1506431.


\end{document}